\documentstyle[12pt,amssymb]{article}
\begin{document}
\title{On the problem of equation of state for the $\Lambda$ - field.}
\author{D. Podolsky\footnote{E-mail: podolsky@itp.ac.ru} \\
Landau Institute for Theoretical Physics\\
2 Kosygin St.,
117334, Moscow, Russia}
\date{18 March 2002}
\maketitle
\begin{abstract}
Recently found accelarated expansion of our Universe is due to the
presence of a new kind of matter called $\Lambda$ - field or quintessence.
The limitations on its equation of state are found from the fact of
its unclustering at all scales much
smaller than the cosmological horizon.
It is discussed how these limitations affect on
the possibility to approximate
the accelarated expansion by such cosmological models as the model of
Chaplygin gas.
\end{abstract}
\newpage
Results of observations of $Ia$ supernovae explosions at high redshifts
are now accepted as strong argument (if not evidence) in the favor of existence
of a new kind of matter playing the role of $\Lambda$ - term in the Einstein
equations (\cite{can1} - \cite{can4}). The question concerning the reconstruction
of lagrangian of such a matter from the observational data then becomes natural.

We have not much information about this kind of matter today: 1) it seems that
cosmological effect due to the  $\Lambda$-field\footnote{the term "quintessence"
was also introduced for a variable $\Lambda$ - term theory \cite{huey}.} is isotropic;
2) the $\Lambda$ - field is practically unclustered (remains spatially homogenious) at all scales
where the distribution of nonrelativistic matter (barions and CDM) is inhomogenious:
$R \le 30 h^{-1} $ Mpc, where $h$ is the Hubble constant in units of $100 $ km/c/Mpc.
The first statement allows us to think that we deal with a scalar mode of some quantum field
theory and its effective largangian takes the following form \cite{bou}:
\begin{equation}
\label{lagr}
L = \frac{1}{2}g^{\mu \nu} \partial_\mu \phi \partial_\nu \phi -
F(\phi ) R - V (\phi )
\end{equation}

It turns out (\cite{as1}, see also \cite{hut},\cite{chi},\cite{bertolami}) that it is possible
to reconstruct the functions $F(\phi )$ and $V(\phi )$ if one has the dependence of
luminocity distance from redshift $D_L (z)$ (for some class of objects,
$Ia$ supernovae, for example) and/or correlation functions (like $\langle \delta (0) \delta (z) \rangle$)
of the dust-like matter density.

Let us remind briefly the basic idea (we will suppose $F(\phi ) = 0$ and follow \cite{as1}).
Cosmological equations for the system with scalar field and dust-like matter have the form:
\begin{equation}
\label{back1}
8\pi G V(\phi ) = aH\frac{dH}{da} + 3H^2 - \frac{3}{2} \Omega_{0,m}
H_0^2 \left( \frac{a_0}{a} \right)^3
\end{equation}
\begin{equation}
\label{back2}
4\pi G a^2 H^2 \left( \frac{d\phi}{da} \right)^2 = -aH \frac{dH}{da} -
\frac{3}{2} \Omega_{0,m} H_0^2 \left( \frac{a_0}{a} \right)^3,
\end{equation}
where $H_0$ is the Hubble parameter for $z=0$, $\Omega_{m,0} = \frac{8\pi G \epsilon_m}{3H_0^2}$
is dimensionless dencity of the dust-like matter for $z = 0$,
$\frac{d}{dt} = aH\frac{d}{da}$.

Once the expression for $H(z)$ is known, it is possible to find $a(\phi )$ from the Eq. (\ref{back2})
and substitute it into Eq. (\ref{back1}) to find the form of potential $V(\phi )$.
The dependence of Hubble parameter on $z$ can be found in turn from the expressions
\begin{equation}
\label{Hz}
H(z) = \left( \frac{d}{dz} \left( \frac{D_L(z)}{1+z} \right) \right)^{-1}
\end{equation}
\begin{equation}
\label{delz}
\frac{H^2(z)}{H^2(0)} = \frac{(1+z)^2 \delta^{'2}(0)}{\delta^{'2}(z)} -
3\Omega_{m,0} \frac{(1+z)^2}{\delta^2(z)} \int_0^z
\frac{\delta |\delta'|}{1 + z} dz
\end{equation}
The eq. (\ref{delz}) is correct only for $F(\phi ) = 0$, otherwise the situation is more
complicated \cite{bou}.

As it is easy to see, the same method works if we would like to find only the equation of state
$p_\Lambda = f(\epsilon_\Lambda)$
without any speculations concerning the form of effective lagrangian for $\Lambda$ - field.
The equations to be solved then have the following form:
\begin{equation}
\label{hyd1}
\frac{8\pi G}{3}\epsilon_\Lambda = H_0^2 \Omega_{m,0} (1 + z)^3 - H^2 (z)
\end{equation}
\begin{equation}
\label{hyd2}
4\pi G f(\epsilon_\Lambda ) = -4\pi G \epsilon_\Lambda -
\frac{3}{2}\Omega_{m,0} H_0^2 (1+z)^3 + H(z) (1 + z) \frac{dH(z)}{dz}
\end{equation}
If $H(z)$ is found, we have to find $z(\epsilon_\Lambda )$ from (\ref{hyd1})
and then substitute it into (\ref{hyd2}).

Modern experimental data correspond to $z \sim 1$ if $H(z)$ is determined from
$D_L (z)$ for supernovae observations (\cite{can1} - \cite{can4}). One
could use the expression (\ref{delz}) to find $H(z)$ from the density of rich
clasters of galaxies $n(z)$ but the data are unsufficiently representative and
exact for now. That is why any information concerning the form of equation
of state will be useful to define the class which the real theory of $\Lambda$ -
term belongs to. Let us find the limitations on the equation of state for $\Lambda$ -
field following from the fact that it is unclustered at all scales much smaller than
the cosmological horizon.

The perturbation theory for the cosmology in which the matter is just the two-component
liquid (non-relativistic matter and $\Lambda$ - field) is determined by the equations
(see, for example, \cite{as2})
\begin{equation}
\label{ptPhi}
\dot{\Phi} + H\Phi = 4\pi G \left( \epsilon_m v_m +
\left( \epsilon_\Lambda + f (\epsilon_\Lambda) \right) v_\Lambda \right)
\end{equation}
\begin{equation}
\label{ptvL}
\frac{d}{dt} \left( \frac{\dot{v}_\Lambda}{f'(\epsilon_\Lambda)} \right)
- 3\frac{d}{dt} \left( Hv_\Lambda \right) + \frac{k^2}{a^2} v_\Lambda =
3 \dot{\Phi} + \frac{d}{dt} \left( \frac{\Phi}{f'(\epsilon_\Lambda )}\right)
\end{equation}
\begin{equation}
\label{ptvm}
\dot{v}_m = \Phi
\end{equation}
\begin{equation}
\label{ptdelta}
\dot{\delta} = \frac{d}{dt} \left( \frac{\delta \epsilon_m}{\epsilon_m}\right) =
12\pi G \left( \epsilon_\Lambda + f (\epsilon_\Lambda )\right)
(v_\Lambda - v_m) - \frac{k^2}{a^2} v_m
\end{equation}
\begin{equation}
\label{ptHvm}
-3 \frac{d}{dt} \left( Hv_m \right) + \frac{k^2}{a^2} v_m = 3\dot{\Phi}
\end{equation}
\begin{equation}
\label{pten}
4\pi G \left( \delta \epsilon_m + \delta \epsilon_\Lambda \right) =
-\frac{k^2}{a^2} \Phi
\end{equation}
\begin{equation}
\label{ptdeL}
\frac{\delta \epsilon_\Lambda}{\epsilon_\Lambda + f(\epsilon_\Lambda )} =
\frac{1}{f'(\epsilon_\Lambda )} \frac{d}{dt} \left( v_\Lambda - v_m  \right),
\end{equation}
where $v$ is the gauge-invariant velocity potential, $\delta$ is the gauge-invariant perturbation
of energy density \cite{bard}, perturbations of FRW metrics are in the longitudinal gauge
($ds^2 = (1+2\Phi) dt^2 - a^2 (t) (1 - 2 \Psi ) \delta_{lm} dx^l dx^m, l,m =  1,2,3$),
Fourier spatial dependence is supposed,
$f'(\epsilon_\Lambda ) = \beta^2_\Lambda = \frac{dp_\Lambda }{d\epsilon_\Lambda }$ and $\Phi = \Psi$
for the given physical system.

Unclustering (in the broad sence) of $\Lambda$ - field at small scales means that the equation
for perturbations of dust-like matter for $r \ll R_H = H^{-1}$ has the same form as it would be
in the case of the absence of $\Lambda$ - field:
\begin{equation}
\label{delta}
\ddot{\delta} + 2H \dot{\delta}  - 4\pi G \epsilon_m \delta = 0
\end{equation}
It can be derived from (\ref{ptPhi}) - (\ref{ptdeL}) as follows. We have from (\ref{ptdelta})
$a^2 \dot{\delta} = 12\pi G (\epsilon_\Lambda + f (\epsilon_\Lambda ))
(v_\Lambda - v_m ) a^2 - k^2 v_m$. Differentiating it with respect to $t$ and using
(\ref{ptvm}), (\ref{ptdeL}) one can find the equation
\begin{equation}
\label{main}
\begin{array}{l}
\ddot{\delta} + 2 H \dot{\delta} = -\frac{k^2}{a^2} \Phi +
12 \pi G (1 + f'(\epsilon_\Lambda))\dot{\epsilon}_\Lambda 
(v_\Lambda - v_m ) + \\
+ 24\pi G H (\epsilon_\Lambda + f(\epsilon_\Lambda ))
(v_\Lambda - v_m ) - 12\pi G f' (\epsilon ) \delta \epsilon_\Lambda
\end{array}
\end{equation}
It is obvious that it cannot be reduced generally to (\ref{delta}) in the case of large $k$
($\frac{k}{aH} \gg 1$). We need to compare the first term in the right-hand side  with
 the others. We have $v_m \sim \frac{a^2}{k^2} \left(  H\Phi \right)$ for short wavelengths  (using
(\ref{ptHvm}) and the condition $\dot{\Phi} \sim H\Phi$). Let us suppose that in the same limit
(\ref{ptvL}) can be reduced to
\begin{equation}
\label{proposal}
v_\Lambda \sim \frac{a^2}{k^2} \left( 3\dot{\Phi} + \frac{d}{dt}
\left(\frac{\Phi}{f'(\epsilon_\Lambda )}\right) \right).
\end{equation}
As it is easy to see, it takes place when
\begin{equation}
\label{c1}
\left(\frac{aH}{k}\right)^2 \left| \frac{1}{f'(\epsilon_\Lambda )} \right| \lesssim 1,
\end{equation}
\begin{equation}
\label{c2}
\left(\frac{aH}{k}\right) \frac{a}{k} \left| \frac{d}{dt}
\left( \frac{1}{f'(\epsilon_\Lambda)} \right) \right| \lesssim 1
\end{equation}
("$x \lesssim 1$" means there that $x$ is not too large with respect to $1$).

The satisfaction of these conditions is sufficient for the first term in the right-hand side of
(\ref{main}) to dominate over the others: let us compare for example the first term with the third term
in the right-hand side of  (\ref{main}):
\begin{equation}
\label{thirdm}
\begin{array}{l}
\pi G (\epsilon_\Lambda + f(\epsilon_\Lambda ))(v_\Lambda - v_m) < \\
< \left( \frac{aH}{k} \right)^2 \Phi
\left( \frac{d}{dt}\left(\frac{1}{f'(\epsilon_m)}\right)
+
\frac{H}{f'(\epsilon_m)} - 3H \right) \ll \frac{k^2}{H a^2}\Phi .
\end{array}
\end{equation}
Similarly it is easy to find that the second term is small (one could use the equation
$\dot{\epsilon_\Lambda} = -3H(\epsilon_\Lambda + f(\epsilon_\Lambda))$).

Let us now show that the inequality $|\delta \epsilon_\Lambda| \ll |\delta \epsilon_m|$
is satisfied for short wavelenghts. It means the unclustering of the $\Lambda$ - term in
the narrow sense. We have from (\ref{ptdeL})
\begin{equation}
\label{nonclust1}
|\delta \epsilon_\Lambda| =
\left| \frac{(\epsilon_\Lambda + f(\epsilon_\Lambda))}{f'(\epsilon_\Lambda)}
\frac{d}{dt}
\left( v_\Lambda - v_m  \right)     \right|
\sim
\frac{H^2}{\pi G} \left( \frac{a}{k}H \right)^2\Phi.
\end{equation}
Using (\ref{pten}) we can find that
$\frac{\delta \epsilon_\Lambda}{\delta \epsilon_m} \sim \left(\frac{aH}{k}\right)^4$.

How strong are the limitations (\ref{c1}), (\ref{c2})? First of all, they can be sensible
only if the QFT describing $\Lambda$-term looks like a hydrodynamics for large scales.
Generally speaking nondiagonal components of the pressure tensor are not small with
respect to diagonal components (i.e., $\sigma_{ik} \ne p\delta_{ik}$) for an arbitrary QFT.
It follows that there is no excitations moving with the speed $\sqrt{\frac{dp}{d\epsilon }}$
or they do not play the key role for a physical system.  So, the hydrodynamical degrees of freedom
are not the physical ones and the answer for the question about the reconstruction of equation
of state is not valuable anymore. In this case we should return to the formulation of inverse
cosmological problem in terms of QFT \cite{as1}.
The language of hydrodynamics is relevant when the free length is small with respect to the
Hubble radius for a corresponding QFT. If we take the theory like (\ref{lagr}) and approximate
the potential energy by
\begin{equation}
\label{decomp}
V(\phi ) \approx V_0 + \frac{m^2 \phi^2}{2} + \frac{\lambda \phi^4}{4} + \cdots ,
\end{equation}
then this condition takes the form
\begin{equation}
\label{lphi4}
l_{kin} \sim \frac{1}{n\sigma} \sim \frac{m}{\Lambda} \frac{m^2}{\lambda^2} \ll \frac{1}{H},
\end{equation}
where $n$ is the density of excitations, $\sigma$ is the cross section in $\phi^4$ - theory, $\Lambda$ is
the energy density of $\Lambda$ - field.

This criterion is rather severe one: if we believe that it is the inflaton field that drives the present
accelerated expansion (and inflation is described by the simplest models with well known
$\lambda \sim 10^{-12}$, $m \sim 10^{13}$ GeV), then the limitations (\ref{c1}), (\ref{c2})
have no relation to reality. Nevertheless, we still have a lot of freedom: a) it may be that the real
inflationary stage which took place in our Universe is described by more complicated model,
multicomponent scalar field, for example; or b) the inflaton has no relation with the present
accelerated expansion.

Let us suppose that the condition (\ref{lphi4}) is satisfied and the hydrodynamics is a good
language for us to describe the dynamics of the $\Lambda$-field. Because of the fact that
$\left( \frac{aH}{k}\right)^2$ is very small (experimental data show the unclustering of $\Lambda$ -term
only for $r < 30$ Mpc), the limitations (\ref{c1}), (\ref{c2}) give no useful information for
the models with the equation of state $p_\Lambda = \omega \epsilon_\Lambda$, where $\omega \le -1/3$.
These models however should be described in terms of QFT but not hydrodynamics: the speed of sound squared
$\frac{dp_\Lambda }{d\epsilon_\Lambda } = \omega$ is negative, it means the exponential growth of
amplitudes of excitations and then the instability of such a hydrodynamics.

The models like Chaplygin gas with the equation of state
$p_\Lambda = - \frac{A}{\epsilon^n_\Lambda}$ (\cite{kamen}, see also \cite{bento})
are more involved from this point  of view. The limitations for $A$ and $n$ following from
(\ref{c1}),(\ref{c2}) turn out to be meaningful:
\begin{equation}
\label{c1c}
\left( \frac{aH}{k} \right)^2 \frac{\Lambda^{n + 1}}{nA} \lesssim 1,
\end{equation}
\begin{equation}
\label{c2c}
\left( \frac{aH}{k} \right)^2 \frac{n + 1}{n A} \Lambda^n \left| \Lambda - \frac{A}{\Lambda^n} \right|
\lesssim 1.
\end{equation}
If the condition $\epsilon_\Lambda + f(\epsilon_\Lambda ) \ge 0$ is satisfied, then
\begin{equation}
\label{n}
(n + 1 ) \left( 1 - \frac{1}{n} \left( \frac{aH}{k}\right)^2 \right) \lesssim 1.
\end{equation}
Thus, $n$ cannot be too large with respect to unity.

I would like to thank A.A. Starobinsky for his useful comments and discussions.

This research is supported by
RFBR, Grants 00-15-96699, 02-02-16817 and the
Fund of Landau Scholarship, Forschungszentrum J{\"u}lich, Germany.


\begin{thebibliography}{99}
\bibitem{can1} S.J. Perlmutter et al., Nature, {\bf 391}, 51 (1998).
\bibitem{can2} P.M. Garnavich et al., Astrophys. J., {\bf 509}, 94 (1998).
\bibitem{can3} A. Riess et al., Astron. J., {\bf 116}, 1009 (1998).
\bibitem{can4} B.P. Schmidt et al., Astrophys. J., {\bf 507}, 46 (1998).
\bibitem{huey} G. Huey, L. Wang, R. Duve, R.R. Caldwell, D.J. Steinhardt, Phys. Rev. D,
{\bf 59}, 063005 (1999).
\bibitem{bou} B. Boisseau, G. Esposito-Farese, D. Polarski,
A. Starobinsky, Phys. Rev. Lett., {\bf 85}, 2236 (2000).
\bibitem{as1} A. Starobinsky, JETP Lett., {\bf 68}, 757 (1998).
\bibitem{hut} D. Huterer, M.S. Turner, Phys. Rev. D, {\bf 60}, 81301 (1999).
\bibitem{chi} T. Nakamura, T. Chiba, MNRAS, {\bf 306}, 696 (1999).
\bibitem{bertolami} O. Bertolami, P.J. Martins, Phys. Rev. D, {\bf 61}, 064007 (2000).
\bibitem{as2} L. Solov'eva, A. Starobinsky, Astron. J., {\bf 62}, 625 (1985) (in russian).
\bibitem{bard} J. Bardeen, Phys. Rev. D, {\bf 22}, 1882 (1980).
\bibitem{kamen} A. Kamenshchik, U. Moschella, V. Pasquier, Phys. Lett. B, {\bf 511}, 265 (2001).
\bibitem{bento} M.C. Bento, D. Bertolami, A.A. Sen, gr-qc/0202064.

\end{thebibliography}
\end{document}